\documentclass[a4paper]{jpconf}
\usepackage{graphicx}
 \usepackage[abs]{overpic}
\usepackage{amsmath}
\usepackage{color}

\begin{document}
\title{Environmental gravitational decoherence\\ and a tensor noise model}

\author{Fumika Suzuki and Friedemann Queisser}

\address{Department of Physics, The University of British Columbia, Vancouver, BC V6T 1Z1, Canada}

\ead{fumika@physics.ubc.ca, queisser@phas.ubc.ca}

\begin{abstract}

We discuss decrease of coherence in a massive system due to the emission of gravitational waves.
In particular we investigate environmental gravitational decoherence in the context of an interference experiment. 
The time-evolution of the reduced density matrix is solved analytically using 
the path-integral formalism.
Furthermore, we study the impact of a tensor noise
onto the coherence properties of massive systems.
We discuss that a particular choice of tensor noise shows similarities to 
a mechanism proposed by Di\'{o}si and Penrose.
\end{abstract}

\section{Introduction}

In recent years, there has been a growing interest in testing gravitational 
decoherence or possible gravitational effects on quantum mechanics (QM) in 
condensed matter and quantum-optical systems \cite{bou, bou2, as}. 
Decoherence can be studied in the framework of quantum mechanics and it does not 
require any additional assumptions.
The dynamics of a system which is coupled with the environment follows from the Schr\"odinger equation.
An observer who has only access to system degrees of freedom observes 
a nonunitary dynamics which can be obtained by tracing out the environmental 
degrees of freedom from the total density matrix.
This averaging generically reduces the coherence of the reduced density
matrix describing the system.
Thus, a part of the phase information is distributed over the environmental variables.
Whether the phase information can be restored or not depends crucially on the form of the coupling and the properties 
of the environment.
We will address decoherence via emission of gravitational waves in section \ref{decoherence}.

Interestingly, it has also been argued that 
gravity might lead to a loss of coherence 
which cannot be discussed in terms of standard quantum 
mechanics \cite{pen,P81,P86,diosi,karo,K74,K69,O92}.
A prominent example of this ``intrinsic decoherence'' 
was discussed in \cite{pen,P81,P86}.
There, it was argued that superpositions of static
configurations have finite life-times and decay on a 
time-scale $T\sim 1/E_g$, where $E_g$ denotes the gravitational 
self-energy of the difference of the mass-distributions
which are involved in the superposition.
The main motivation for such an effect is the lack of a
canonical time-like killing vector when superpositions of
space-times are considered.
In section \ref{noise}, we want to address the question whether
it is possible to derive such a decoherence rate from a tensor noise model.
We point out several problems which come along when one introduces a tensor noise ``by hand''.

\section{Environmental gravitational decoherence}\label{decoherence}

Every physical system is coupled to gravity via its energy-momentum tensor.
In the weak-field limit, the metric $g_{\mu\nu}$ can be expanded 
around the Minkowski background $\eta_{\mu\nu}$ according to $g_{\mu\nu}=\eta_{\mu\nu}+h_{\mu\nu} $.
The resulting action is quadratic in the metric perturbations $h_{\mu\nu}$.
Using the path integral formalism, the influence functional can be evaluated 
exactly.
When the metric perturbations have a time-dependent quadrupole moment, the 
environmental modes can carry away phase-information about the system which 
will lead to a real loss decoherence \cite{bill}.
%
%
In the following, we discuss the loss of coherence of a particle which does not 
move on a geodesic and radiates gravitational waves.
The analogue effect in electrodynamics due to emission 
of photons (bremsstrahlung) has been studied in \cite{bre, bre2}.

\subsection{Influence functional}

In this section, we follow notations used in \cite{ana}. The total action of a mass distribution which is centered around $\mathbf{x}_0$ 
and coupled to gravity can be written as
\begin{equation}
S=S_{\rm sys} (\mathbf{x}_0)+S_{\rm grav} (h_{\mu\nu} )+S_{\rm int} (h_{\mu\nu}, \mathbf{x}_0).
\end{equation}

The action $S_{\rm grav} (h_{\mu\nu} )$ contains in general also gravitational self-interaction terms
which are of higher order and will not be considered here.
We choose the transverse-traceless gauge, i.e., $h_{0\mu}=h^{ij}_{\hspace{.2cm},i}=h^i_{\hspace{.1cm}i}=0$. 
Then the graviton action adopts the simple quadratic form
\begin{equation}
S_{\rm grav} (h_{\mu\nu}) = \frac{1}{4\pi G}\int d^4 x h_{\mu\nu ,\alpha}h^{\mu\nu ,\alpha}=\frac{1}{4\pi G} \int d^4 x (\partial_{t} h_{ij}\partial^{t}h^{ij}-h_{ij, k}h^{ij, k}).
\end{equation}

The interaction between the energy-momentum tensor $T^{ij}_{\mathbf{x}_0}$ of the
system and the external graviton field is bilinear, i.e.,
\begin{equation}
S_{\rm int} (h_{\mu\nu} , \mathbf{x}_0) =\int d^4 x h_{ij}T^{ij}_{\mathbf{x}_0}.
\end{equation}

Here only the spatial components of the energy-momentum tensor contribute due to the gauge choice above.
The graviton field can be expanded into plane waves according to
\begin{equation}
h_{ij} (\mathbf{x}, t) =\int \frac{d^3 k}{(2\pi)^3}\displaystyle\sum_{\lambda} \epsilon^{\lambda}_{ij} 
(\mathbf{k}) \left( q^{\lambda}_{s} (t) \cos (\mathbf{k}\cdot \mathbf{x}) +q^{\lambda}_{a} (t) \sin (\mathbf{k}\cdot \mathbf{x})\right)
\end{equation}
where we introduced the transverse-traceless polarization tensors $\epsilon^{\lambda}_{ij} (\mathbf{k})$. 
The time-evolution of the system's density matrix is determined by the 
action $S_\mathrm{sys}(\mathbf{x}_0)$ and the influence functional $\mathcal{F}(\mathbf{x}_0,\mathbf{x}_0')$.
Then the propagator of the reduced density matrix is given by
\begin{equation}
W (\mathbf{x}_{0, i}, \mathbf{x}_{0,f} ; \mathbf{x}'_{0,i}, \mathbf{x}'_{0,f})=
\int_{\mathbf{x}_{0,i}}^{\mathbf{x}_{0,f}} D\mathbf{x}_0 \int^{\mathbf{x}'_{0,f}}_{\mathbf{x}'_{0,i}}
D\mathbf{x}'_0 \exp (iS_{\rm sys} (\mathbf{x}_0)-i S_{\rm sys} (\mathbf{x}_0'))\mathcal{F} (\mathbf{x}_0 , \mathbf{x}_0')
\end{equation}
where the subscripts $i, f$ indicate the initial and final values respectively. 
The influence functional adopts the explicit form
\begin{eqnarray}\label{influence}
\mathcal{F} (\mathbf{x}_0 , \mathbf{x}_0')=\exp (i \phi) &=& \exp \Biggr[ i\int_0^{T} d^4 x
\int^{t}_0 d^4 x' (T^{ij}_{\mathbf{x}_0} (x) +T^{ij}_{\mathbf{x}'_0} (x) )\gamma_{ij ,kl} (x-x')
(T^{kl}_{\mathbf{x}_0} (x') -T^{kl}_{\mathbf{x}'_0} (x') )\nonumber\\
&&-\int_0^{T} d^4 x \int^{t}_0 d^4 x' (T^{ij}_{\mathbf{x}_0} (x) -T^{ij}_{\mathbf{x}'_0} (x) )
\eta_{ij ,kl} (x-x') (T^{kl}_{\mathbf{x}_0} (x') -T^{kl}_{\mathbf{x}'_0} (x') )\Biggr]\nonumber\\
\end{eqnarray}
with the correlation functions
\begin{eqnarray}
\gamma_{ij, kl} (x-x') &=& \frac{G}{8\pi^2}\int \frac{d^3 k}{|\mathbf{k}|}\sin (|\mathbf{k}| (t-t'))
e^{i\mathbf{k} \cdot (\mathbf{x}-\mathbf{x}')}\Pi_{ij, kl} (\mathbf{k}),\nonumber\\
\eta_{ij, kl} (x-x') &=& \frac{G}{8\pi^2}\int \frac{d^3 k}{|\mathbf{k}|}\cos (|\mathbf{k}| (t-t'))
\coth\left(\frac{\beta |\mathbf{k}|}{2}\right)e^{i\mathbf{k} \cdot (\mathbf{x}-\mathbf{x}')}\Pi_{ij, kl} (\mathbf{k}).
\end{eqnarray}

We assumed the environmental modes to be in a thermodynamical state with the 
inverse temperature $\beta$.
The sum over polarizations gives the polarization tensor $\sum_{\lambda} \epsilon^{\lambda}_{ij} (\mathbf{k}) \epsilon^{\lambda *}_{kl} 
(\mathbf{k}) =\Pi_{ij , kl} (\mathbf{k}) = P_{ik}P_{jl}+P_{il}P_{jk}-P_{ij}P_{kl}$ where the projection operators are $P_{ij}=\delta_{ij}-\frac{k_{i}k_{j}}{|\mathbf{k}|^2}$. 

\subsection{Time-dependent behaviour of decoherence functional}

\begin{figure}[t]
 \begin{minipage}{14pc}
  \begin{overpic}[width=14pc]{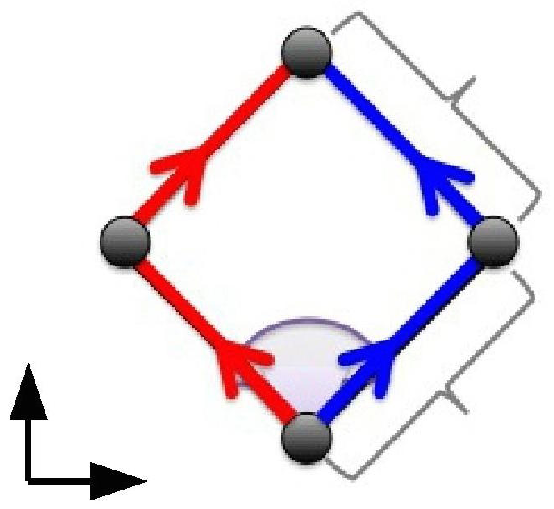}
 \put(50,45){$v$}
 \put(50,110){$v$}
 \put(117,110){$v$}
 \put(133,124){$a$}
 \put(117,45){$v$}
 \put(85,40){$\theta$}
 \put(75,3){$t=0$}
 \put(45,10){$\hat{\mathbf{x}}$}
 \put(8,48){$\hat{\mathbf{y}}$}
 \put(-15,76){$t_{\rm kick}=\frac{T}{2}$}
 \put(147,76){$t_{\rm kick}=\frac{T}{2}$}
 \put(75,143){$t=T$}
 \put(130,26){$a$}
 \end{overpic}
 \caption{\label{setup}Typical trajectories followed by a system in a quantum superposition experiment \cite{bre, bre2}. In this diagram, $t=t_{\rm kick}=\frac{T}{2}$.}
 \end{minipage}\hspace{2pc}%
 \begin{minipage}{22pc}
 \includegraphics[width=22pc]{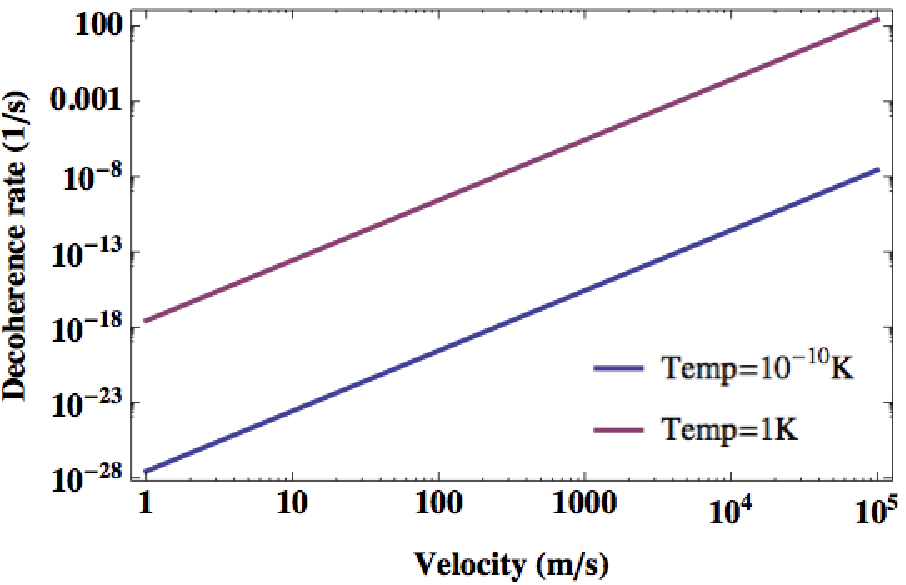}
 \caption{\label{label}The dependence of decoherence rate on velocity $v$ of an object
with temperature of gravitational waves $T=1$K and $T=10^{-10}$K.
We chose the mass of an object $m=10^{-5}$kg.}
 \end{minipage} 
 \end{figure}

The real and imaginary parts of the phase $\phi$ in the influence functional (\ref{influence}) lead to dissipation and decoherence, respectively. 
In this section, we calculate
\begin{equation}\label{phase}
 \mbox{Im}(\phi)=-\int_0^{T} d^4 x \int^{t}_0 d^4 x' (T^{ij}_{\mathbf{x}_0} (x) -T^{ij}_{\mathbf{x}'_0} (x) )
 \eta_{ij ,kl} (x-x') (T^{kl}_{\mathbf{x}_0} (x') -T^{kl}_{\mathbf{x}'_0} (x') )
\end{equation}
from which the dependence of the decoherence rate on the various parameters of the model follows.
As particular example, we discuss the evaluation of (\ref{phase}) using an interference device setup (Figure~\ref{setup}).
A matter distribution is in superposition of either following a right trajectory $\mathbf{x}_0$ (blue) or a left one $\mathbf{x}'_0$ (red).
For a small angle $\theta$, Figure~\ref{setup}  can be interpreted as the double-slit experiment, whereas $\theta=\frac{\pi}{2}$ corresponds to an interferometer.
On both trajectories, the system changes its direction at time $t=t_{\rm kick}$ due to a slit or mirror.
The energy-momentum tensor of a particle in flat space-time takes the 
form
\begin{eqnarray}
T_{\mathrm{part}, \mathbf{x}_0}^{\mu\nu}(x)= \frac{m v^\mu(t) v^\nu(t)}{\sqrt{1-|\mathbf{v}_0(t)|^2}}\delta_\sigma (\mathbf{x}-\mathbf{x}_0(t))
\end{eqnarray}
with the velocity vector $v_\mu(t)=(1,\mathbf{v}_0(t))$.
We assume that a particle is smeared out over a space-region 
determined by $\sigma$, i.e.,
$
\delta_\sigma(\mathbf{x})=\frac{1}{(2\pi \sigma^2)^{3/2}}\exp\left[-\frac{\mathbf{x}^2}{2\sigma^2}\right]\,.
$

It is well-known that the divergence equation of the energy-momentum tensor
gives the geodesic equation with respect to the background space-time.
However, the two possible trajectories of a particle in Figure~\ref{setup} do 
not correspond to geodesics in Minkowski space-time, hence $\partial_\nu T_{\mathrm{part},\mathbf{x}_0}^{\mu\nu}\neq 0$.
Generally, one would need to model a mirror, the interaction between
a particle and a mirror and the interaction of gravity with the particle-mirror 
system.
Here, we follow a simpler approach by adding momentum densities 
which kick a particle at $t=t_{\rm kick}$.
Thus, our system can be described now as ``particle + momentum density such 
that the energy-momentum tensor is divergence-free along the trajectories''.
In particular, we choose 
\begin{eqnarray}
T_{\mathrm{mom},\mathbf{x}_0}^{01}(x)=-m \int^t_{-\infty}ds \delta_\sigma (\mathbf{x}-\mathbf{x}_0(s))
\frac{d}{ds}\left(\frac{v^1(s)}{\sqrt{1-|\mathbf{v}_0(s)|^2}}\right),
\\
T_{\mathrm{mom},\mathbf{x}_0}^{02}(x)=-m \int^t_{-\infty}ds\delta_\sigma (\mathbf{x}-\mathbf{x}_0(s))
\frac{d}{ds}\left(\frac{v^2(s)}{\sqrt{1-|\mathbf{v}_0(s)|^2}}\right).
\end{eqnarray}

The time-time component $T_{\mathrm{mom},\mathbf{x}_0}^{00}(x)$ can be chosen such that
\begin{eqnarray}
\partial_\alpha ( T_{\mathrm{mom},\mathbf{x}_0}^{\alpha 0}(x)+T_{\mathrm{part},\mathbf{x}_0}^{\alpha 0}(x))=0.
\end{eqnarray}

The additional momentum densities ensure that the divergence of the energy-momentum tensor 
$T_{\mathbf{x}_0}^{\mu\nu} =T_{\mathrm{part},\mathbf{x}_0}^{\mu\nu}+T_{\mathrm{mom},\mathbf{x}_0}^{\mu\nu}$
vanishes along the trajectory.

In order to obtain an interpretation of the momentum densities introduced,
let us consider $\mathbf{v}_0(s)=v_1(s)\hat{\mathbf{x}}+v_2\hat{\mathbf{y}}$ with 
$v_1(s)=v\sin(\theta/2)\theta(t_{\rm kick}-s)-v\sin(\theta/2) \theta(s-t_{\rm kick})$ and $v_2=v\cos(\theta/2)$. 
Then we have
\begin{eqnarray}
-m \int^t_{-\infty}ds \delta_\sigma (\mathbf{x}-\mathbf{x}_0(s))
\frac{d}{ds}\left(\frac{v^1}{\sqrt{1-|\mathbf{v}_0(s)|^2}}\right)=-\frac{2mv\sin(\theta/2)}{\sqrt{1-v^2}}\theta(t-t_0)\delta_\sigma (\mathbf{x}-\mathbf{x}_0(t_0))\,.
\end{eqnarray}

We interpret this expression as follows:
When $t<t_{\rm kick}$, a particle moves straight in positive $\hat{\mathbf{x}}$- and $\hat{\mathbf{y}}$-directions.
At time $t_{\rm kick}$, it gets a momentum transfer which changes its
velocity in $\hat{\mathbf{x}}$-direction from $v\sin(\theta/2)$ to $-v\sin(\theta/2)$. 
This corresponds to the right trajectory $\mathbf{x}_0$ in Figure~\ref{setup}. 
A left trajectory $\mathbf{x}_0'$ which has a kick at $t_{\rm kick}$ towards the right can be constructed in the same way by defining
$\mathbf{v}'_0(s)=-v_1(s)\hat{\mathbf{x}}+v_2\hat{\mathbf{y}}$.

By substituting $T_{\mathbf{x}_0}^{\mu\nu} =T_{\mathrm{part},\mathbf{x}_0}^{\mu\nu}+T_{\mathrm{mom},\mathbf{x}_0}^{\mu\nu}$ 
into (\ref{phase}) and evaluating the thermal part of the expression (\ref{phase}) along the trajectories, we find for velocities $v\ll1$
\begin{eqnarray}
\mathrm{Im}(\phi)_\mathrm{therm}=&\frac{4Gm^2v^4}{105\pi}\left[42(1+\cos(\theta))+(63+64\cos(\theta)+5\cos(2\theta))v^2\right]
\sin^2\left(\frac{\theta}{2}\right)\nonumber\\
&\times\left(4\ln\left[\frac{\sinh\left(\pi T/(2\beta\right))}{\pi T/(2\beta)}\right]
-\ln\left[\frac{\sinh\left(\pi T/\beta\right)}{\pi T\beta}\right]\right)+\mathcal{O}(v^8)
\end{eqnarray}
where $T$ is a time when two trajectories meet again.
With an angle $0<\theta<\pi$, the effect is approximately proportional to $v^4$.
The authors of \cite{hu} 
found the same velocity-dependence in the non-relativistic single-mode limit 
from a master equation approach.
When $\pi T/(2\beta)\gg 1$ and $\theta=\pi/2$, we have
\begin{equation}
\mathrm{Im}(\phi)_\mathrm{therm}=\frac{4 Gm^2 T v^4}{5 \beta}.
\end{equation}

The decoherence factor $\mathrm{Im}(\phi)$ vanishes when two trajectories are parallel $\theta=0$, that is,
the trajectories coincide with each other.
With $\theta=\pi$, $\mathrm{Im}(\phi)$ is proportional to $v^6$,
\begin{equation}
\mathrm{Im}(\phi)_\mathrm{therm}=\frac{16 Gm^2 T v^6}{105 \beta}\,. 
\end{equation}

Apart from thermal gravitational waves, vacuum fluctuations also contribute to the 
decoherence rate.
For realistic parameters, this contribution is very small that it can be 
neglected.
Its expression diverges logarithmically with the inverse of the 
spread of the mass distribution, $1/\sigma$.

In Figure 2, we plotted the decoherence rate with the temperature of the gravitational waves being 1K and $10^{-10}$K. Since gravitational waves decoupled
earlier from photons and have therefore a lower temperature than the Cosmic Microwave background.
However, the actual temperature of the thermal
gravitational waves background can be much lower and it depends on the details of reheating after inflation \cite{hu2}. 
Here, we assume
a thermal contribution with 1K as the upper bound and vacuum contribution would
be the lower bound on the decoherence rate. 

\subsection{Gravitational radiation}

We claim that the decoherence effect comes along with the emission of gravitational waves.
The emitted energy of a particle along one of the trajectories 
depicted in Figure~\ref{setup} can be calculated classically.
The emission occurs when the direction of a particle is changed by 
a mirror.
As long as the emitted radiation can be distinguished for trajectories, the environment measures which path a particle 
is moving to some extent, and this leads to environmental gravitational decoherence.

For a trace-reversed perturbation, 
$\bar{h}_{\mu\nu}^\mathrm{pert}=\bar{h}_{\mu\nu}^\mathrm{pert}-\bar{h}^\mathrm{pert}\eta_{\mu\nu}/2$,
the solution to the linearized Einstein equations gives the expression
\begin{align}
\bar{h}_{\mu\nu}^\mathrm{pert}(t,\mathbf{x})=4 G\int d^3 y \frac{T_{\mu\nu}(t-|\mathbf{x}-\mathbf{y}|,\mathbf{y})}
{|\mathbf{x}-\mathbf{y}|}\, .
\end{align}

When the particle is slowly moving, the far field is dominated by 
quadrupole moment tensor $I_{ij}=\int d^3x\, x^i x^j T_{00,\mathbf{x}_0}(x)$.
In terms of the trace-free quadrupole moment, $J_{ij}=I_{ij}-\delta_{ij}\delta^{kl}I_{kl}/3$,
the emitted power is
\begin{align}
P(t,r)=-\frac{G}{5}\frac{d^3 J_{ij}(t-r)}{dt^3}\frac{d^3 J^{ij}(t-r)}{dt^3} .
\end{align}

Here we will assume a smooth trajectory of a particle which moves along 
one path
\begin{align}
\mathbf{v}_0(t)=v\sin(\theta/2)[f(t,0)-2f(t,T/2)+f(t,T)]\hat{\mathbf{x}}+
v\cos(\theta/2)[f(t,0)-f(t,T)]\hat{\mathbf{y}}\,
\end{align}
with $f(t,s)=(\tanh((t-s)/t_0)-1)/2$.
The time $t_0$ determines how fast the direction of a particle is changed at a mirror, how fast a 
particle is accelerated initially from a source and decelerated at a detector finally.
In the limit $T\gg t_0$, the radiated energy is crucially determined 
by $t_0$,
\begin{align}
E\sim -\frac{v^4G m^2}{t_0^5}\,. 
\end{align}

For vacuum fluctuations, the minimal possible time $t_0$ is determined by the 
extension of the particle, that is $t_{0,\mathrm{min}}\sim \sigma$.
For finite temperatures, the effective cutoff of the gravitational modes is determined
by $1/\beta$, so one should expect $t_{0,\mathrm{min}}\sim \beta$.
%

\section{Tensor noise model}\label{noise}

It has been argued that the incompatibility between general relativity and quantum mechanics 
may lead to a form of intrinsic decoherence \cite{pen,P81,P86}.
Such an effect, if it exists, is not related to any form of environmental degrees of freedom 
which carry away phase information from the system.
The conflict between the principle of 
general covariance and the superposition 
principle of quantum mechanics is fundamental:
When gravity is dynamical, the geometry of space-time must depend on the
quantum state of the matter which it contains.   
Thus, different matter states live in different
space-times which cannot be identified pointwise with each 
other.
One might argue that the difference of three-forces in the superposed 
space-times give an estimate for the typical life-time of a 
superposition.
This life-time $t=1/\Gamma$ is given by $\sim 1/E_g$ with
\begin{align}\label{gamma_penrose}
E_g&=\int d^3\mathbf{x}(\mathbf{f}(x)-\mathbf{f'}(x))\cdot(\mathbf{f}(x)-\mathbf{f'}(x))\nonumber\\
  &=-4\pi G \int d\mathbf{x}d\mathbf{x}'\frac{[\rho(\mathbf{x})-\rho'(\mathbf{x})]
[\rho(\mathbf{x}')-\rho'(\mathbf{x}')]}{|\mathbf{x}-\mathbf{x'}|}.
\end{align}

Although the relation (\ref{gamma_penrose}) has not been derived from 
fundamental physics and although it is not clear yet, whether an intrinsic decoherence exists in nature at all \cite{hu3}, we want to
address the question, whether this decoherence rate can be derived in some limit from a 
noise model. 
Scalar noise models have been studied in the literature, see, for example \cite{gordon}.
However, we want to couple the energy-momentum tensor to the noise 
which requires a tensor random current.
Furthermore, we assume that the life-time (\ref{gamma_penrose}) is the Newtonian limit of a more general expression.
A possible generalization would be
\begin{align}\label{gamma_penrose_mod}
\Gamma\sim &\int dt' \int d\mathbf{x}\int d\mathbf{x'} T_{\mu\nu}(\mathbf{x},t)D^{\mu\nu,\lambda\rho}(\mathbf{x}-\mathbf{x}',t-t')T_{\lambda\rho}(\mathbf{x'},t')
\end{align}
where $D_{\mu\nu,\lambda\rho}$ is the Green's function for linearized gravitational waves.
Since $D_{\mu\nu,\lambda\rho}$ is not positive 
definite, we are immediately confronted with 
a problem: The expression (\ref{gamma_penrose_mod}) can 
also adopt negative values, which invalidates the interpretation
of a decoherence rate.
In the context of quantum mechanics, this would 
violate the positivity of the density matrix 
since the off-diagonal elements would be unbounded,
\begin{align}
\langle i |\hat{\rho}(T)|j\rangle\approx \exp\left[-\int_0^T dt \Gamma(t)\right]
\langle i |\hat{\rho}(0)|j\rangle, \qquad i\neq j\,. 
\end{align}

However, it might be possible that terms such as (\ref{gamma_penrose}) or (\ref{gamma_penrose_mod}) are only 
a part of a more general \textit{positive} definite quantity.
The simplest covariant terms 
which can be added to the Fierz-Pauli action take the form
\begin{align}\label{gravnoise}
S_j=\int d^4x j^{\mu\nu}\left(a\,h_{\mu\nu}+b\, T_{\mu\nu}\right)\,
\end{align}
with arbitrary constants $a$ and $b$.
Here we introduced the tensor noise $j_{\mu\nu}$ which couples to 
the linearized gravitational field and to the energy-momentum tensor.
The first term would correspond to a stochastic contribution 
of the energy-momentum tensor whereas the second part 
resembles stochastic gravitational fluctuations.
The gauge-invariance of the graviton field requires $\partial_\mu j^{\mu\nu}=0$.
Furthermore we assume the Gaussian distribution $\mathcal{P}$
of the tensor noise to be
\begin{align}
\mathcal{P}[j_{\mu\nu}]=\frac{1}{N}\exp
\left(-\frac{1}{2}\int d^4x j_{\mu\nu}
j^{\mu\nu}\right)\,.
\end{align}

Contrary to section \ref{decoherence}, we will use the de-Donder gauge.
This includes also the instantaneous interaction between 
stationary matter distributions in addition to the radiation part 
of the gravitational perturbations.
After performing the integrals over $j_{\mu\nu}$ and $h_{\mu\nu}$, 
additional contributions to the influence functional arise due to the 
presence of the noise.
In the non-relativistic limit, we find for large times 
the expression
\begin{align}
\mathcal{F}= \mathcal{F}_0\times\exp\left[-(\Gamma_1+\Gamma_2+\Gamma_3)T\right]\,,
\end{align}
where $\mathcal{F}_0$ denotes the influence functional due to 
gravitational waves.
Assuming an energy-momentum tensor of the form $T_{\mu\nu}=\rho u_\mu u_\nu$,
the decoherence rates $\Gamma_i$ have a simple interpretation.
$\Gamma_1$ measures the difference of the
Newtonian gravitational potentials in the respective 
spacetimes,
\begin{align}\label{gamma1}
\Gamma_1=\frac{\kappa^2a^2}{8\pi^2}\int d^3\mathbf{x}(\Phi(\mathbf{x})-\Phi'(\mathbf{x}))^2\,.
\end{align}

$\Gamma_2$ is equal  
to the expression (\ref{gamma_penrose})
up to a numerical factor and 
accounts for the difference of the Newtonian forces,
\begin{align}\label{gamma2}
\Gamma_2=-\frac{2ab}{\kappa}\int d^3\mathbf{x}(\mathbf{f}(\mathbf{x})-\mathbf{f}'(\mathbf{x}))\cdot
(\mathbf{f}(\mathbf{x})-\mathbf{f}'(\mathbf{x}))\,.
\end{align}

Dimensional arguments might suggest that $ab\sim\kappa$.
Finally, $\Gamma_3$ incorporates the difference of
the mass densities,
\begin{align}\label{gamma3}
\Gamma_3=\frac{b^2}{2}\int d^3\mathbf{x}(\rho(\mathbf{x})-\rho'(\mathbf{x}))^2\,.
\end{align}

It is not yet clear whether this result is a 
gauge-independent statement.
The time-evolution of the density matrix requires a definite choice 
of the time-parameter $t$.
(We are not interested in scattering matrix elements where
the interaction decreases adiabatically to zero for $t\rightarrow\pm \infty$ 
and Lorentz-invariant quantities can be defined.)
A change in the time-parameter, corresponding to an transformation 
of the metric components, might alter the expression for the decoherence rate.

\section{Conclusion}

We studied the decrease of coherence due to the emission 
of gravitational waves and gave an estimate for the 
emitted gravitational energy.
The dominant contribution of the decoherence rate 
is given by thermal gravitational waves.
Due to the smallness of the gravitational interaction, this 
effect is negligible for elementary particles.
If it would be possible to send an object with mass $m=10^{-6}$kg
and velocity $v=10^{5}$m/s through an interferometer, the decoherence 
rate would be roughly $\Gamma= 3$s$^{-1}$ with temperature of gravitational waves being 1K.

Furthermore, we discussed a tensor noise model which 
partly resembles, in the Newtonian limit, a form of intrinsic decoherence 
which was discussed in the literature before.

The coherence decrease due to this noise model depends on the difference of the mass
density, the gravitational potential and the Newtonian 3-force 
of the superposition.

\ack

 The authors would like to thank G.~W.~Semenoff, D.~Carney, D.~Scott, B.~L.~Hu, 
 C.~Anastopoulos and W.~G.~Unruh for helpful discussions, and express our gratitude to 
 P.~C.~E.~Stamp for support. F.~S.~is partially supported by a UBC international tuition award, 
 UBC faculty of science graduate award, the research foundation for opto-science and technology, NSERC Discovery grant and CFI funds for CRUCS. 
 The part of the work was supported by the Templeton foundation (grant number JTF 36838).


\section*{References}
\begin{thebibliography}{9}
\bibitem{bou} Marshall W, Simon C, Penrose R and Bouwmeester D 2003 {\it Phys. Rev. Lett.} {\bf 91} 130401
\bibitem{bou2} Kleckner D et al. 2008  {\it New Journal of Physics.} {\bf 10} 095020 
\bibitem{as} Kaltenbeck R et al. 2012 {\it Exp Astron.} {\bf 34} 123
\bibitem{pen} Penrose R 1996 {\it General Relativity and Gravitation.} {\bf 28} 581-599
\bibitem{P81} Penrose R 1986, In \textit{Quantum Gravity 2: A Second Oxford Symposium}, Oxford University Press
\bibitem{P86} Penrose R 1986, In \textit{Quantum Concepts in Space and Time}, Oxford University Press
\bibitem{diosi}  Di\'{o}si L 1987 {\it Phys. Lett.} {\bf 120} A 377-381
\bibitem{karo} K\'{a}rolyh\'{a}zy F 1966 {\it Nuovo Cim.} {\bf 52} 390
\bibitem{K74} K\'arolyh\'azy F 1974, Magyar Fizikai Polyoirat {\bf 12} 24
\bibitem{K69} Komar A B 1969, Int.~J.~Theor.~Phys.~{\bf 2} 157
\bibitem{O92} Omn\`es R 1992, Rev.~Mod.~Phys.~{\bf 64} 339
\bibitem{bill} Unruh W G 2000 {\it In Relativistic quantum measurement and decoherence} eds Breuer H P and Petruccione F (Springer) pp 125-140
\bibitem{bre} Breuer H and Petruccione F 2001 {\it Phys. Rev.} A {\bf 63} 032102-1
\bibitem{bre2} Breuer H and Petruccione F 2002 {\it The Theory of Open Quantum Systems.} (Oxford)
\bibitem{ana} Anastopoulos C 1996 {\it Phys. Rev.} D {\bf 54} 1600
\bibitem{hu}  Anastopoulos C and Hu B L 2013 {\it Class. Quantum Grav.} {\bf 30} 165007
\bibitem{hu2} Hu B L 2014 {\it J. Phys.: Conf. Ser.} {\bf 504} 012021 
\bibitem{hu3}  Anastopoulos C and Hu B L 2008 {\it Class. Quantum Grav.} {\bf 25} 154003
\bibitem{gordon} Cloutier J and Semenoff G W 1991 {\it Phys. Rev.} D {\bf 44} 3218-3229








\end{thebibliography}
\end{document}